\documentclass[aps,prl,twocolumn,epsfig,superscriptaddress,floatfix]{revtex4}
\usepackage{graphicx}
\usepackage{bm}
\usepackage{color}
\usepackage{times}

\begin{document}
\title{Exciton binding energies in  carbon nanotubes from two-photon photoluminescence}
\hyphenation{na-no-tubes Ryd-berg}

\author{J. Maultzsch}
\email{janina@physik.tu-berlin.de} \affiliation{Institut f\"ur Festk\"orperphysik, Technische Universit\"at
Berlin, Hardenbergstr. 36, 10623 Berlin, Germany}

\author{R. Pomraenke}
\affiliation{Max-Born-Institut f\"ur Nichtlineare Optik und Kurzzeitspektroskopie, 12489 Berlin, Germany}

\author{S. Reich}
\affiliation{Department of Engineering, University of Cambridge, Cambridge CB2 1PZ, United Kingdom}

\author{E. Chang}
\affiliation{INFM National Research Center S3, and Physics Department, University of Modena and Reggio Emilia, 41100 Modena, Italy}

\author{D. Prezzi}
\affiliation{INFM National Research Center S3, and Physics Department, University of Modena and Reggio Emilia, 41100 Modena, Italy}

\author{A. Ruini}
\affiliation{INFM National Research Center S3, and Physics Department, University of Modena and Reggio Emilia, 41100 Modena, Italy}

\author{E. Molinari}
\affiliation{INFM National Research Center S3, and Physics Department, University of Modena and Reggio Emilia, 41100 Modena, Italy}

\author{M. S. Strano}
\affiliation{University of Illinois, Deptartment of Chemistry and Biomolecular Engineering, Urbana, IL 61801,
USA}

\author{C. Thomsen}
\affiliation{Institut f\"ur Festk\"orperphysik, Technische Universit\"at Berlin, Hardenbergstr. 36, 10623
Berlin, Germany}

\author{C. Lienau}
\affiliation{Max-Born-Institut f\"ur Nichtlineare Optik und Kurzzeitspektroskopie, 12489 Berlin, Germany}
\begin{abstract}
One- and two-photon luminescence excitation spectroscopy showed a series of distinct excitonic states in
single-walled carbon nanotubes. The energy splitting between one- and two-photon-active
exciton states of different wavefunction symmetry is the  fingerprint of excitonic interactions in carbon nanotubes. We determine exciton binding energies of $0.3-0.4$\,eV for different nanotubes with diameters
between 6.8\,\AA{} and 9.0\,\AA. Our results, which are supported by \emph{ab-initio} calculations of the linear and non-linear optical spectra, prove that the elementary optical excitations of carbon nanotubes are strongly Coulomb-correlated, quasi-one dimensionally confined electron-hole pairs, stable even at room temperature. This alters our  microscopic understanding of both the electronic structure and the Coulomb interactions in carbon nanotubes, and has direct impact on the optical and transport properties of novel nanotube devices.
\end{abstract}
\maketitle

Single-walled carbon nanotubes are fascinating nano-objects~\cite{reich04}. Their unique geometric,
electronic and optical properties hold promise for a variety of novel applications, including nanoscale field-effect transistors
, electrically excited single-molecule light sources, 
and nanosensing~\cite{Tans98,Misewich03,Snow05}. Initially, their optical properties have received comparatively little attention,
as photoluminescence is quenched in nanotube bundles by non-radiative relaxation of excited carriers
via metallic tubes. This has changed fundamentally since the discovery of band-gap luminescence from individual tubes in stable aqueous solutions~\cite{oconnell02}, enabling direct studies of their optical
properties~\cite{bachilo02,hartschuh03b}.

 Nanotube optical spectra have mostly been interpreted in terms of single-particle excitations so far~\cite{bachilo02}, governed by the van-Hove singularities in the density of states of a quasi-one-dimensional system. 
Recent theoretical studies, in contrast, predict that  excitonic effects, \emph{i.e.}, the Coulomb interaction between the  excited electron and hole, affect both the transition frequencies and  the shape of the optical spectra~\cite{Ando97,chang04,spataru04,perebeinos04,zhao04,kane04}.
Because of
the strong quantum confinement of electron and hole in the quasi-one-dimensional nanotube, large exciton binding energies are  expected, but calculated values span a  broad range from a few tens of meV~\cite{perebeinos04} to
1\,eV~\cite{spataru04,chang04}. 
It is of utmost importance to reconcile
these conflicting points of view. 
First, the strength of the Coulomb correlation directly influences the energetic ordering of electronic states and thus the correlation between optical and structural properties. Also, the existence of optically inactive (``dark") excitonic states may cause the unusually low luminescence quantum yield~\cite{zhao04}  in nanotubes. Finally, the transport properties of nanotube devices are affected as optical excitations may either move as uncorrelated electrons and holes or as bound excitons~\cite{freitag04}.

So far, experimental evidence for excitonic correlations in nanotubes is indirect. Optical absorption and
emission energies differ from the single-particle predictions (``ratio problem")~\cite{bachilo02}. Also, ultrafast
intersubband relaxation times are thought to support the exciton picture~\cite{korovyanko04}. Yet,  direct
experimental proof is lacking and, in particular, the exciton binding energies are unknown.  
In this Letter, we directly identify exciton binding effects in
single-walled carbon nanotubes by one- and two-photon luminescence. We determine exciton binding energies of 300 to 400\,meV. Supported by \emph{ab initio} calculations, our results show that the elementary optical excitations of carbon nanotubes are Coulomb-correlated electron-hole pairs that are stable even at room temperature.

A classical example for two-particle Coulomb correlations is the  Rydberg series in atomic hydrogen
spectroscopy. The same energy series also governs the bound excitonic states $E^n=E_{\mathrm{gap}} - E^{\mathrm R}/n^2$ in homogeneous three-dimensional bulk semiconductors ($E_{\mathrm{gap}}$: band gap energy, $E^{\mathrm R}$: 3D exciton Rydberg energy).  
In carbon nanotubes, on the other hand, the electron density is confined to the plane of the
rolled graphite sheet. Excitonic wavefunctions are expected to be delocalized along the circumference of the tube and to extend over several nanometers along the tube axis~\cite{chang04}. 

\begin{figure}[t]
\begin{center}\includegraphics[width=8.0cm,clip]{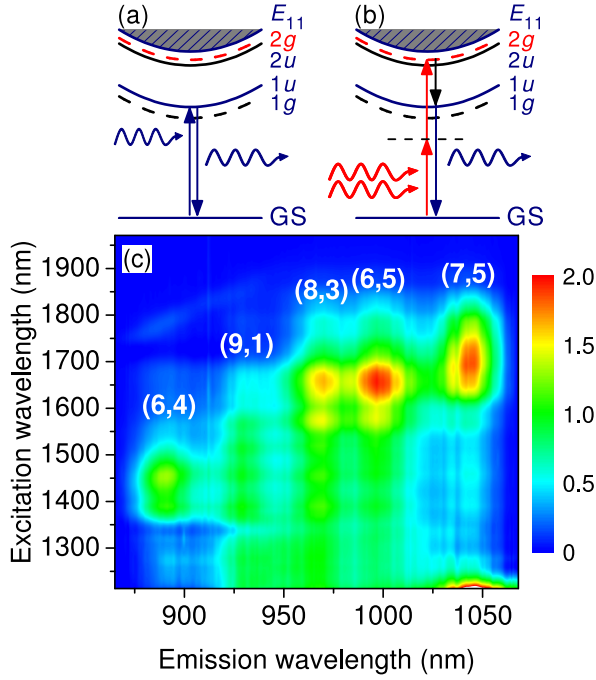}  \end{center}
\vspace{-0.5cm} \caption{\emph{(color online)} (a) Schematic picture of one-photon absorption and emission in single-walled carbon nanotubes. $E_{11}$ indicates the single-particle transition between the lowest subbands. One-photon excitations
couple to excitonic states with 
odd ($u$) symmetry with respect to $\pi$ rotations about the $U$ axis.  
The $U$ axis is perpendicular to the tube axis through the center of one of the C hexagons~\cite{damnjanovic99a}. $(1)$ and $(2)$ indicate the symmetry of the envelope function with respect to reflection in the $z=0$ plane, see also Fig.~\ref{fig3b}. 
Emission occurs from the lowest one-photon active $1u$ state.  (b) Two-photon absorption results in the excitation of
exciton states with even ($g$) symmetry under the $U$-axis operation.  (c) Two-photon  luminescence spectra of carbon nanotubes. The
 luminescence intensity is plotted as a function of excitation and detection wavelength. Each emission
peak comes from a different nanotube species; the chiral index is indicated.  \label{fig1}}
\end{figure}

The concept of our experiment is to validate the excitonic character of optical excitations in carbon nanotubes by
addressing excitonic states with different wave function symmetry, analogous to the $s, p, d,...$ states in the hydrogen atom. Briefly, in one-photon spectroscopy, one
expects for each allowed interband transition with single-particle gap $E_{\mathrm{gap}}$ a series of
transitions to excitonic states with odd ($u$) symmetry with respect to rotations by $\pi$ about the $U$ axis of
the nanotube [Fig.~\ref{fig1}\,(a)]. Two-photon spectroscopy, on the other hand, couples to the otherwise optically inactive even ($g$) states [Fig.~\ref{fig1}\,(b)]. The energetic splitting between the one- and two-photon observed states indicates the strength of Coulomb  correlations. Therefore, two-photon absorption is an elegant technique to determine excitonic effects~\cite{Rinaldi94}.

\begin{figure}[t]
\begin{center}\includegraphics[width=8.0cm,clip]{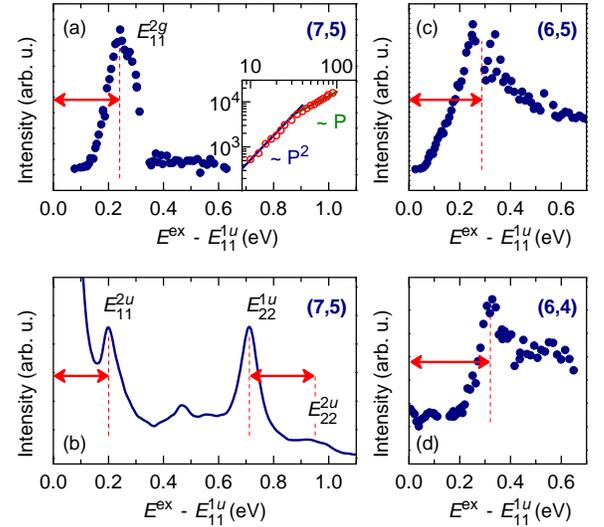}  \end{center}
\vspace{-0.5cm} \caption{\emph{(color online)} (a) Two-photon photoluminescence intensity of the (7,5) tube (at $ 1045\pm5$\,nm) shown
as a function of excitation energy $E^{\mathrm{ex}}$. The abscissa gives the difference between two-photon
absorption and emission energy. The $2g$ exciton state is excited at $0.24$\,eV above the $1u$ resonance.
Inset: Excitation-power (in mW) dependence of luminescence intensity. (b) Same as in (a) but for one-photon
absorption. $E_{11}$ and $E_{22}$ indicate the first- and second-subband transition, respectively. The
peak at $0.2$\,eV is assigned to the $2u$ exciton resonance, consistent with observation of the $2g$ 
state at $0.24$\,eV. (c) and (d) Two-photon photoluminescence excitation spectra of the (6,5) and the (6,4) tube, respectively.} \label{fig2}
\end{figure}

We investigated single-walled carbon nanotubes suspended in D$_2$O wrapped by a surfactant~\cite{oconnell02}. 
The sample was excited at room temperature with 150\,fs pulses from an optical parametric oscillator between 1150\,nm and 2000\,nm with an average power of 60\,mW. The luminescence was recorded in $90^\circ$ configuration by a charge-coupled device. In the  region 
between 850\,nm and 1070\,nm [Fig.~\ref{fig1}\,(c)], we identified the luminescence from 5 different nanotube species [(6,4), (9,1), (8,3), (6,5), and (7,5)]. We also observed the weak emission from the (9,4) tube at 1130 nm~\cite{fussnote_redshift}. The assignment of the chiral indices $(n_1,n_2)$ is based on  Raman~\cite{telg04} and luminescence~\cite{bachilo02} data.

Figure~\ref{fig1}\,(c) shows the  photoluminescence for excitation below the band gap at wavelengths between 1210 and 1970\,nm. 
For each tube
we find a  pronounced maximum in the luminescence intensity at an excitation wavelength far above the
emission wavelength, but significantly smaller than twice this wavelength. We assign these absorption maxima to
resonant two-photon excitation of the lowest two-photon allowed exciton state ($2g$). The energetic positions of the
maxima thus correspond to half the energy of this state. Emission results from  relaxation
 into the lowest one-photon active $1u$ state [Fig.~\ref{fig1}\,(b)]. The shift of about $240 - 320$\,meV
between the energies of both exciton states is a  signature of (i) the excitonic nature of absorption and
emission at room temperature  and (ii)  exciton binding energies in carbon nanotubes of almost one fourth of the
band-gap energy.

An analyis of the power dependence of the emission intensity  supports the assignment to two-photon
absorption, see inset to Fig.~\ref{fig2}\,(a). At low excitation powers, we observe a quadratic increase of the two-photon photoluminescence. This saturates into a linear increase at powers above 40 mW.

To quantitatively analyze the splitting between the one- and two-photon allowed transitions, we plot two-photon photoluminescence excitation spectra for different nanotube species [Fig.~\ref{fig2} (a), (c), and (d)]. 
These curves
show the emission intensity as a function of the energy difference $E^{\mathrm{ex}} - E_{11}^{1u}$ between
two-photon excitation and emission energy. 
For the (7,5) tube [Fig.~\ref{fig2}\,(a)],  the resonance
maximum is at 240\,meV above the one-photon active state. For the (6,5) and (6,4) tubes [Fig.~\ref{fig2} (c) and (d)], the $E_{11}^{2g} - E_{11}^{1u}$ splittings are slightly larger, 285 and 325 meV, respectively. 

From the energy splitting $E_{11}^{2g} - E_{11}^{1u}$, we estimate  the exciton binding energy
by assuming as a  first approximation the  two-dimensional (2D) hydrogen model. In this model, the   exciton binding energies are given by $E^{\mathrm{b}}=E_{\mathrm{gap}}-E^{1u}=9/8\,(E^{2g}-E^{1u})$, see Table~\ref{tab1}. If the exciton is delocalized along the circumference of the tube~\cite{pedersen03},  a better description is given by a  variational model of an electron and hole moving under an attractive Coulomb interaction  on a
cylindrical surface with dielectric screening $\epsilon$~\cite{fussnote_cylinder}. With the diameter of tube given, $\epsilon$ is the only adjustable parameter and is  found from the experimental energy difference $E_{11}^{2g} - E_{11}^{1u}$.
With $\epsilon \approx 12.5$, we thus obtain a binding energy $E^{\mathrm {b}}=0.31$\,eV for the (7,5) tube and even larger values for the other investigated species (Table~\ref{tab1}).

Based on the hydrogenic model, we expect to observe a second one-photon allowed absorption resonance at an
energy close to the two-photon active state. In Fig.~\ref{fig2}\,(b),  we plot a one-photon photoluminescence
excitation spectrum of the (7,5) tube.  The spectrum shows an absorption resonance at an excess energy of 200\,meV, \emph{i.e.},  slightly below the first two-photon resonance at 240\,meV. The same resonance is also
observed in the excitation spectrum of the second subband ($E_{22}$).  These resonances  hence reflect
one-photon excitation of higher-lying exciton states. This supports a previously proposed assignment of
one-photon excitation spectra~\cite{reich05b_sub}, whereas alternative suggestions for the origin of these peaks, including phonon sidebands and exciton-phonon excitations~\cite{perebeinos05,qiu05}, are  less likely.

\begin{figure}[t]
\vspace{-0.5cm} \begin{center}\includegraphics[width=8.0cm,clip]{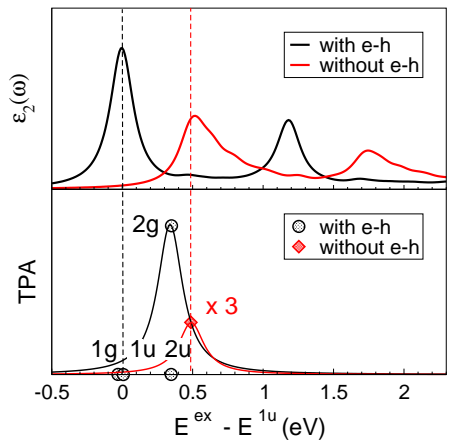}  \end{center} 
\caption{ \emph{(color online)}
{\it Ab initio} calculated one- and two- photon absorption (top and bottom panel,
respectively) for the (6,4) nanotube.
Black and red lines are with and without electron-hole (e-h) 
interactions, respectively.
 A Lorentzian broadening of  0.1 eV has been 
introduced
everywhere. In the bottom panel, the four black circles 
correspond to the probability amplitude of two-photon scattering to the four exciton
states $1g$, $1u$, $2u$, and $2g$; the red diamond shows the probability
amplitude of scattering to the final active state without e-h 
interaction. Higher-energy exciton states with zero or negligible amplitude
are omitted for clarity. The horizontal axis is the photon energy relative
to the lowest optically active exciton ($1u$).
} \label{fig3}
\end{figure}

\begin{figure}[t]
\begin{center}\includegraphics[width=8.0cm,clip]{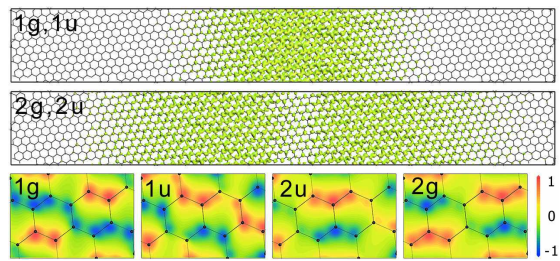}  \end{center}
\caption{ \emph{(color online)}
Lowest-energy excitonic wavefunctions for the (6,4) nanotube
as obtained from {\it ab initio} calculations. The top two panels
correspond to the $1g$, $1u$ and $2g$, $2u$ states,  respectively, and show
the probability of finding the electron on the tube surface when the hole
is fixed at the center of the panel ($z=0$). The vertical direction corresponds to
the circumference  (2.1 nm) and the horizontal direction to 
the tube axis (each panel is 15.9 nm long).
The bottom four panels are blow-ups of the same states, 
showing the wavefunction symmetry. They display the wavefunction
amplitude of the electron when the hole is placed in the center
of the bond at $z=0$ and show 
the parity of the wavefunctions under rotation by $\pi$ about the $U$ axis.
 The color scale  is in arbitrary units.
}
\label{fig3b}
\end{figure}

For a more thourough understanding of the excitonic states in carbon nanotubes, 
we perform first-principles calculations of the excited states starting from  single-particle wavefunctions described by  density functional theory (DFT) in the GW approximation. This method 
involves the solution of an effective, two-particle,
Schr\"{o}dinger-like equation (the Bethe-Salpeter Equation, BSE) describing the dynamics of the electron and
hole under their mutual Coulomb (attractive) interaction. It therefore 
permits a calculation of the
wavefunctions and binding energies of bound electron-hole pairs~\cite{benedict98,albrecht98,rohlfing98,ruini02}. 

We  calculate the two-photon absorption intensity
\emph{vs.} the energy of the final excited state in the (6,4) tube (Fig.~\ref{fig3}). Within this atomistic description, the two
inequivalent C atoms of each hexagon lead to a doubling of the states with respect to the cylindrical hydrogenic
model, the states in each doublet being very close in energy. We find four states below the single-particle gap
(red line), corresponding to four bound exciton states, with binding energies of 0.54 ($1g$), 0.50 ($1u$), 0.16
($2g$) and 0.16 eV ($2u$). Their parities under  $\pi$-rotation about the $U$ axis are consistent with the fact that the $u$ states are one-photon active and the $g$ states are two-photon allowed. The envelope functions of the strongly bound states are approximately symmetric ($1$) under reflection in the $z=0$ plane; the weakly bound states have an approximately anti-symmetric ($2$) envelope.
Our calculations predict that the most strongly bound exciton state is optically inactive in one-photon absorption and weak in two-photon absorption~\cite{zhao04}. This may explain the small luminescence quantum yield in carbon nanotubes observed even at low temperatures.
\begin{table*}
\begin{ruledtabular}
\begin{tabular}{ccccccc}
$(n_1,n_2)$ & $d$ (\AA ) & $E_{11}^{1u}$ (eV) & $E_{11}^{2g}$ (eV)& $E_{11}^{2g}-E_{11}^{1u}$ (eV) &
[$E_{11}^{\mathrm {b}}]_a$ (eV)& [$E_{11}^{\mathrm {b}}]_b$ (eV)\\\hline
(6,4) & 6.83 & 1.395 & 1.720 & 0.325 & 0.365 & 0.42 \\
(9,1) & 7.47 & 1.335 & 1.650 & 0.315 & 0.355 & 0.42 \\
(8,3) & 7.72 & 1.275 & 1.570 & 0.295 & 0.330 & 0.38 \\
(6,5) & 7.47 & 1.245 & 1.530 & 0.285 & 0.320 & 0.37 \\
(7,5) & 8.18 & 1.190 & 1.430 & 0.240 & 0.270 & 0.31 \\
(9,4) & 9.03 & 1.095 & 1.375 & 0.280 & 0.315 & 0.38 
\end{tabular}
\end{ruledtabular}
\caption{Observed nanotube structures with their diameter $d$, their emission energy ($E_{11}^{1u}$) and the
two-photon absorption energy ($E_{11}^{2g}$). The binding energy is estimated within two models: the
two-dimensional hydrogen model gives $[E^{\mathrm {b}}_{11}]_a $; the cylindrical hydrogen model gives
$[E^{\mathrm {b}}_{11}]_b$. In the latter model, different values for $\epsilon$ ranging from 9 to 11 were
taken for each tube to match the experimentally measured values of $E^{2g}-E^{1u}$.}\label{tab1}
\end{table*}

If  we neglect the electron-hole interaction, the two-photon absorption is very close in energy to the first one-photon active state (red curves in Fig.~\ref{fig3}). 
On the other hand, if the electron-hole interaction is included, these two energies ($E^{2g}_{11}$ and $E^{1u}_{11}$) differ by 0.34\,eV, which  
agrees well  with the measured value of 0.325\,eV, see Table~\ref{tab1}. We calculate a binding energy of 0.5\,eV from  first-principles. Similarly, a binding energy of 0.33\,eV is found for the (8,4) tube.
The lowest exciton wavefunction  extends over several nanometers along the tube axis and is 
 delocalized along the nanotube circumference. The higher exciton states ($2g$, $2u$) are  more extended along the tube axis and have a nodal plane at $z=0$, see Fig.~\ref{fig3b}.
Consequently, the  calculated excitonic wavefunctions are  Wannier-like and weakly dependent on  the circumference direction.
Furthermore, the calculated band structure  confirms that a two-band model is appropriate, see Ref.~\cite{chang04}. The \emph{ab-initio} approach thus provides a complete knowledge of the excitonic states, 
and  validates \emph{a posteriori} the cylindrical hydrogen model.

Based on the hydrogen model as well as on previous theoretical work~\cite{kane04,zhao04,perebeinos04}, we expect an increase of the  exciton binding energy with decreasing tube diameter. The energy should also depend on the
family index~\cite{reich00c} $\nu=(n_1-n_2)\,\textrm{mod }3=\pm1$, where larger binding energies $E^{\mathrm {b}}_{11}$ are expected for $\nu=-1$ tubes than for $\nu=+1$ tubes with similar diameter~\cite{perebeinos04,zhao04}.
Our experimental results indeed show an overall decrease for larger tube diameters (Table~\ref{tab1}).
All tubes in our experiment have $\nu=-1$, except for the (6,5) tube. The (6,5) tube has the same diameter as the (9,1) tube, but  a lower binding energy. 
 Our data  support the predicted
trends~\cite{chang04,spataru04,perebeinos04,zhao04,kane04} on both the diameter and family-index  dependence; a final conclusion may be drawn once a larger number of tube species has been studied.

In conclusion, we  experimentally observed  a series of exciton states with different wavefunction symmetry in carbon nanotubes by one- and two-photon luminescence. We determined
 exciton binding energies between 300 and 400\,meV for
several nanotube species, in good agreement with \emph{ab initio} calculations. Our results 
directly demonstrate that excitonic properties dominate
the optical absorption and emission in carbon nanotubes even at room temperature.

This work was supported in part by the DFG (SFB296), by an INFM Supercomputing grant at Cineca (Italy), and by the Engineering and Physical Sciences Research Council (EPSRC). S.R. was supported by the Oppenheimer Fund and Newnham College.


\begin{thebibliography}{30}
\expandafter\ifx\csname natexlab\endcsname\relax\def\natexlab#1{#1}\fi
\expandafter\ifx\csname bibnamefont\endcsname\relax
  \def\bibnamefont#1{#1}\fi
\expandafter\ifx\csname bibfnamefont\endcsname\relax
  \def\bibfnamefont#1{#1}\fi
\expandafter\ifx\csname citenamefont\endcsname\relax
  \def\citenamefont#1{#1}\fi
\expandafter\ifx\csname url\endcsname\relax
  \def\url#1{\texttt{#1}}\fi
\expandafter\ifx\csname urlprefix\endcsname\relax\def\urlprefix{URL }\fi
\providecommand{\bibinfo}[2]{#2}
\providecommand{\eprint}[2][]{\url{#2}}

\bibitem[{\citenamefont{Reich et~al.}(2004)\citenamefont{Reich, Thomsen, and
  Maultzsch}}]{reich04}
\bibinfo{author}{\bibfnamefont{S.}~\bibnamefont{Reich}},
  \bibinfo{author}{\bibfnamefont{C.}~\bibnamefont{Thomsen}}, \bibnamefont{and}
  \bibinfo{author}{\bibfnamefont{J.}~\bibnamefont{Maultzsch}},
  \emph{\bibinfo{title}{Carbon {N}anotubes: {B}asic {C}oncepts and {P}hysical
  {P}roperties}} (\bibinfo{publisher}{Wiley-VCH}, \bibinfo{address}{Berlin},
  \bibinfo{year}{2004}).

\bibitem[{\citenamefont{Tans et~al.}(1998)\citenamefont{Tans, Verschueren, and
  Dekker}}]{Tans98}
\bibinfo{author}{\bibfnamefont{S.~J.} \bibnamefont{Tans}},
  \bibinfo{author}{\bibfnamefont{A.}~\bibnamefont{Verschueren}},
  \bibnamefont{and} \bibinfo{author}{\bibfnamefont{C.}~\bibnamefont{Dekker}},
  \bibinfo{journal}{Nature} \textbf{\bibinfo{volume}{393}},
  \bibinfo{pages}{6680} (\bibinfo{year}{1998}).

\bibitem[{\citenamefont{Misewich et~al.}(2003)\citenamefont{Misewich, Martel,
  Tsang, Heinze, and Tersoff}}]{Misewich03}
\bibinfo{author}{\bibfnamefont{J.~A.} \bibnamefont{Misewich}},
  \bibinfo{author}{\bibfnamefont{R.}~\bibnamefont{Martel}},
  \bibinfo{author}{\bibfnamefont{J.~C.} \bibnamefont{Tsang}},
  \bibinfo{author}{\bibfnamefont{S.}~\bibnamefont{Heinze}}, \bibnamefont{and}
  \bibinfo{author}{\bibfnamefont{J.}~\bibnamefont{Tersoff}},
   \bibinfo{journal}{Science} \textbf{\bibinfo{volume}{300}},
  \bibinfo{pages}{783} (\bibinfo{year}{2003}).

\bibitem[{\citenamefont{Snow et~al.}(2005)\citenamefont{Snow, Perkins, Houser,
  Badescu, and Reinecke}}]{Snow05}
\bibinfo{author}{\bibfnamefont{E.~S.} \bibnamefont{Snow}},
  \bibinfo{author}{\bibfnamefont{F.~K.} \bibnamefont{Perkins}},
  \bibinfo{author}{\bibfnamefont{E.~J.} \bibnamefont{Houser}},
  \bibinfo{author}{\bibfnamefont{S.~C.} \bibnamefont{Badescu}},
  \bibnamefont{and} \bibinfo{author}{\bibfnamefont{T.~L.}
  \bibnamefont{Reinecke}}, 
  \bibinfo{journal}{Science}
  \textbf{\bibinfo{volume}{307}}, \bibinfo{pages}{1942} (\bibinfo{year}{2005}).

\bibitem[{\citenamefont{{O'C}onnell et~al.}(2002)\citenamefont{{O'C}onnell,
  Bachilo, Huffman, Moore, Strano, Haroz, Rialon, Boul, Noon, Kittrell
  et~al.}}]{oconnell02}
\bibinfo{author}{\bibfnamefont{M.~J.} \bibnamefont{{O'C}onnell}},
  \bibinfo{author}{\bibfnamefont{S.~M.} \bibnamefont{Bachilo}},
  \bibinfo{author}{\bibfnamefont{C.~B.} \bibnamefont{Huffman}},
  \bibinfo{author}{\bibfnamefont{V.~C.} \bibnamefont{Moore}},
  \bibinfo{author}{\bibfnamefont{M.~S.} \bibnamefont{Strano}},
  \bibinfo{author}{\bibfnamefont{E.~H.} \bibnamefont{Haroz}},
  \bibinfo{author}{\bibfnamefont{K.~L.} \bibnamefont{Rialon}},
  \bibinfo{author}{\bibfnamefont{P.~J.} \bibnamefont{Boul}},
  \bibinfo{author}{\bibfnamefont{W.~H.} \bibnamefont{Noon}},
  \bibinfo{author}{\bibfnamefont{C.}~\bibnamefont{Kittrell}},
   \bibinfo{journal}{Science}
  \textbf{\bibinfo{volume}{297}}, \bibinfo{pages}{593} (\bibinfo{year}{2002}).

\bibitem[{\citenamefont{Bachilo et~al.}(2002)\citenamefont{Bachilo, Strano,
  Kittrell, Hauge, Smalley, and Weisman}}]{bachilo02}
\bibinfo{author}{\bibfnamefont{S.~M.} \bibnamefont{Bachilo}},
  \bibinfo{author}{\bibfnamefont{M.~S.} \bibnamefont{Strano}},
  \bibinfo{author}{\bibfnamefont{C.}~\bibnamefont{Kittrell}},
  \bibinfo{author}{\bibfnamefont{R.~H.} \bibnamefont{Hauge}},
  \bibinfo{author}{\bibfnamefont{R.~E.} \bibnamefont{Smalley}},
  \bibnamefont{and} \bibinfo{author}{\bibfnamefont{R.~B.}
  \bibnamefont{Weisman}}, \bibinfo{journal}{Science}
  \textbf{\bibinfo{volume}{298}}, \bibinfo{pages}{2361} (\bibinfo{year}{2002}).

\bibitem[{\citenamefont{Hartschuh et~al.}(2003)\citenamefont{Hartschuh,
  Pedrosa, Novotny, and Krauss}}]{hartschuh03b}
\bibinfo{author}{\bibfnamefont{A.}~\bibnamefont{Hartschuh}},
  \bibinfo{author}{\bibfnamefont{H.~N.} \bibnamefont{Pedrosa}},
  \bibinfo{author}{\bibfnamefont{L.}~\bibnamefont{Novotny}}, \bibnamefont{and}
  \bibinfo{author}{\bibfnamefont{T.~D.} \bibnamefont{Krauss}},
  \bibinfo{journal}{Science} \textbf{\bibinfo{volume}{301}},
  \bibinfo{pages}{1354} (\bibinfo{year}{2003}).


\bibitem[{\citenamefont{Ando}(1997)}]{Ando97}
\bibinfo{author}{\bibfnamefont{T.}~\bibnamefont{Ando}}, \bibinfo{journal}{J.
  Phys. Soc. Jpn.} \textbf{\bibinfo{volume}{66}}, \bibinfo{pages}{1066}
  (\bibinfo{year}{1997}).

\bibitem[{\citenamefont{Chang et~al.}(2004)\citenamefont{Chang, Bussi, Ruini,
  and Molinari}}]{chang04}
\bibinfo{author}{\bibfnamefont{E.}~\bibnamefont{Chang}},
  \bibinfo{author}{\bibfnamefont{G.}~\bibnamefont{Bussi}},
  \bibinfo{author}{\bibfnamefont{A.}~\bibnamefont{Ruini}}, \bibnamefont{and}
  \bibinfo{author}{\bibfnamefont{E.}~\bibnamefont{Molinari}},
  \bibinfo{journal}{Phys. Rev. Lett.} \textbf{\bibinfo{volume}{92}},
  \bibinfo{pages}{196401} (\bibinfo{year}{2004}).

\bibitem[{\citenamefont{Spataru et~al.}(2004)\citenamefont{Spataru,
  Ismail-Beigi, Benedict, and Louie}}]{spataru04}
\bibinfo{author}{\bibfnamefont{C.~D.} \bibnamefont{Spataru}},
  \bibinfo{author}{\bibfnamefont{S.}~\bibnamefont{Ismail-Beigi}},
  \bibinfo{author}{\bibfnamefont{L.~X.} \bibnamefont{Benedict}},
  \bibnamefont{and} \bibinfo{author}{\bibfnamefont{S.~G.} \bibnamefont{Louie}},
  \bibinfo{journal}{Phys. Rev. Lett.} \textbf{\bibinfo{volume}{92}},
  \bibinfo{pages}{077402} (\bibinfo{year}{2004}).

\bibitem[{\citenamefont{Perebeinos et~al.}(2004)\citenamefont{Perebeinos,
  Tersoff, and Avouris}}]{perebeinos04}
\bibinfo{author}{\bibfnamefont{V.}~\bibnamefont{Perebeinos}},
  \bibinfo{author}{\bibfnamefont{J.}~\bibnamefont{Tersoff}}, \bibnamefont{and}
  \bibinfo{author}{\bibfnamefont{P.}~\bibnamefont{Avouris}},
  \bibinfo{journal}{Phys. Rev. Lett.} \textbf{\bibinfo{volume}{92}},
  \bibinfo{pages}{257402} (\bibinfo{year}{2004}).

\bibitem[{\citenamefont{Zhao and Mazumdar}(2004)}]{zhao04}
\bibinfo{author}{\bibfnamefont{H.}~\bibnamefont{Zhao}} \bibnamefont{and}
  \bibinfo{author}{\bibfnamefont{S.}~\bibnamefont{Mazumdar}},
  \bibinfo{journal}{Phys. Rev. Lett.} \textbf{\bibinfo{volume}{93}},
  \bibinfo{pages}{157402} (\bibinfo{year}{2004}).

\bibitem[{\citenamefont{Kane and Mele}(2004)}]{kane04}
\bibinfo{author}{\bibfnamefont{C.~L.} \bibnamefont{Kane}} \bibnamefont{and}
  \bibinfo{author}{\bibfnamefont{E.~J.} \bibnamefont{Mele}},
  \bibinfo{journal}{Phys. Rev. Lett.} \textbf{\bibinfo{volume}{93}},
  \bibinfo{pages}{197402} (\bibinfo{year}{2004}).

\bibitem[{\citenamefont{Freitag et~al.}(2004)\citenamefont{Freitag, Chen,
  Tersoff, Tsang, Fu, Lie, and Avouris}}]{freitag04}
\bibinfo{author}{\bibfnamefont{M.}~\bibnamefont{Freitag}},
  \bibinfo{author}{\bibfnamefont{J.}~\bibnamefont{Chen}},
  \bibinfo{author}{\bibfnamefont{J.}~\bibnamefont{Tersoff}},
  \bibinfo{author}{\bibfnamefont{J.~C.} \bibnamefont{Tsang}},
  \bibinfo{author}{\bibfnamefont{Q.}~\bibnamefont{Fu}},
  \bibinfo{author}{\bibfnamefont{J.}~\bibnamefont{Lie}}, \bibnamefont{and}
  \bibinfo{author}{\bibfnamefont{P.}~\bibnamefont{Avouris}},
  \bibinfo{journal}{Phys. Rev. Lett.} \textbf{\bibinfo{volume}{93}},
  \bibinfo{pages}{076803} (\bibinfo{year}{2004}).

\bibitem[{\citenamefont{Korovyanko et~al.}(2004)\citenamefont{Korovyanko,
  Sheng, Vardeny, Dalton, and Baughman}}]{korovyanko04}
\bibinfo{author}{\bibfnamefont{O.~J.} \bibnamefont{Korovyanko}},
  \bibinfo{author}{\bibfnamefont{C.-X.} \bibnamefont{Sheng}},
  \bibinfo{author}{\bibfnamefont{Z.}~\bibnamefont{Vardeny}},
  \bibinfo{author}{\bibfnamefont{A.~B.} \bibnamefont{Dalton}},
  \bibnamefont{and} \bibinfo{author}{\bibfnamefont{R.~H.}
  \bibnamefont{Baughman}}, \bibinfo{journal}{Phys. Rev. Lett.}
  \textbf{\bibinfo{volume}{92}}, \bibinfo{pages}{017403}
  (\bibinfo{year}{2004}).

\bibitem[{\citenamefont{Rinaldi et~al.}(1994)\citenamefont{Rinaldi, Cingolani,
  Lepore, Ferrara, Catalano, Rossi, Rota, Molinari, and Lugli}}]{Rinaldi94}
\bibinfo{author}{\bibfnamefont{R.}~\bibnamefont{Rinaldi}},
  \bibinfo{author}{\bibfnamefont{R.}~\bibnamefont{Cingolani}},
  \bibinfo{author}{\bibfnamefont{M.}~\bibnamefont{Lepore}},
  \bibinfo{author}{\bibfnamefont{M.}~\bibnamefont{Ferrara}},
  \bibinfo{author}{\bibfnamefont{I.~M.} \bibnamefont{Catalano}},
  \bibinfo{author}{\bibfnamefont{F.}~\bibnamefont{Rossi}},
  \bibinfo{author}{\bibfnamefont{L.}~\bibnamefont{Rota}},
  \bibinfo{author}{\bibfnamefont{E.}~\bibnamefont{Molinari}}, \bibnamefont{and}
  \bibinfo{author}{\bibfnamefont{P.}~\bibnamefont{Lugli}},
  \bibinfo{journal}{Phys. Rev. Lett.} \textbf{\bibinfo{volume}{73}},
  \bibinfo{pages}{2899} (\bibinfo{year}{1994}).

\bibitem[{\citenamefont{Damnjanovi{\a'c}
  et~al.}(1999)\citenamefont{Damnjanovi{\a'c}, Milo{\v{s}}evi{\a'c},
  Vukovi{\'c}, and Sre{\-}da{\-}no{\-}vi{\'c}}}]{damnjanovic99a}
\bibinfo{author}{\bibfnamefont{M.}~\bibnamefont{Damnjanovi{\a'c}}},
  \bibinfo{author}{\bibfnamefont{I.}~\bibnamefont{Milo{\v{s}}evi{\a'c}}},
  \bibinfo{author}{\bibfnamefont{T.}~\bibnamefont{Vukovi{\'c}}},
  \bibnamefont{and}
  \bibinfo{author}{\bibfnamefont{R.}~\bibnamefont{Sre{\-}da{\-}no{\-}vi{\'c}}},
  \bibinfo{journal}{Phys. Rev. B} \textbf{\bibinfo{volume}{60}},
  \bibinfo{pages}{2728} (\bibinfo{year}{1999}).

\bibitem[{fus({\natexlab{a}})}]{fussnote_redshift}
\bibinfo{note}{The emission energies are about $25$\,meV smaller than reported
  by Bachilo \emph{et al.}~\cite{bachilo02}, which is probably due to the
  presence of small bundles in our sample.}

\bibitem[{\citenamefont{Telg et~al.}(2004)\citenamefont{Telg, Maultzsch, Reich,
  Hennrich, and Thomsen}}]{telg04}
\bibinfo{author}{\bibfnamefont{H.}~\bibnamefont{Telg}},
  \bibinfo{author}{\bibfnamefont{J.}~\bibnamefont{Maultzsch}},
  \bibinfo{author}{\bibfnamefont{S.}~\bibnamefont{Reich}},
  \bibinfo{author}{\bibfnamefont{F.}~\bibnamefont{Hennrich}}, \bibnamefont{and}
  \bibinfo{author}{\bibfnamefont{C.}~\bibnamefont{Thomsen}},
  \bibinfo{journal}{Phys. Rev. Lett.} \textbf{\bibinfo{volume}{93}},
  \bibinfo{pages}{177401} (\bibinfo{year}{2004}).

\bibitem[{\citenamefont{Pedersen}(2003)}]{pedersen03}
\bibinfo{author}{\bibfnamefont{T.~G.} \bibnamefont{Pedersen}},
  \bibinfo{journal}{Phys. Rev. B} \textbf{\bibinfo{volume}{67}},
  \bibinfo{pages}{073401} (\bibinfo{year}{2003}).

\bibitem[{fus({\natexlab{b}})}]{fussnote_cylinder}
\bibinfo{note}{See EPAPS document no. xxx for a description of the
  cylinder model.}

\bibitem[{\citenamefont{Reich et~al.}(manuscript in
  preparation)\citenamefont{Reich, Maultzsch, and Thomsen}}]{reich05b_sub}
\bibinfo{author}{\bibfnamefont{S.}~\bibnamefont{Reich}},
  \bibinfo{author}{\bibfnamefont{J.}~\bibnamefont{Maultzsch}},
  \bibnamefont{and} \bibinfo{author}{\bibfnamefont{C.}~\bibnamefont{Thomsen}}
  (\bibinfo{year}{manuscript in preparation}).

\bibitem[{\citenamefont{Perebeinos et~al.}(2005)\citenamefont{Perebeinos,
  Tersoff, and Avouris}}]{perebeinos05}
\bibinfo{author}{\bibfnamefont{V.}~\bibnamefont{Perebeinos}},
  \bibinfo{author}{\bibfnamefont{J.}~\bibnamefont{Tersoff}}, \bibnamefont{and}
  \bibinfo{author}{\bibfnamefont{P.}~\bibnamefont{Avouris}},
  \bibinfo{journal}{Phys. Rev. Lett.} \textbf{\bibinfo{volume}{94}},
  \bibinfo{pages}{027402} (\bibinfo{year}{2005}).

\bibitem[{\citenamefont{Qiu et~al.}(2005)\citenamefont{Qiu, Freitag,
  Perebeinos, and Avouris}}]{qiu05}
\bibinfo{author}{\bibfnamefont{X.}~\bibnamefont{Qiu}},
  \bibinfo{author}{\bibfnamefont{M.}~\bibnamefont{Freitag}},
  \bibinfo{author}{\bibfnamefont{V.}~\bibnamefont{Perebeinos}},
  \bibnamefont{and} \bibinfo{author}{\bibfnamefont{P.}~\bibnamefont{Avouris}},
  \bibinfo{journal}{Nano Lett.}  (\bibinfo{year}{2005}).

\bibitem[{\citenamefont{Benedict et~al.}(1998)\citenamefont{Benedict, Shirley,
  and Bohn}}]{benedict98}
\bibinfo{author}{\bibfnamefont{L.}~\bibnamefont{Benedict}},
  \bibinfo{author}{\bibfnamefont{E.}~\bibnamefont{Shirley}}, \bibnamefont{and}
  \bibinfo{author}{\bibfnamefont{R.}~\bibnamefont{Bohn}},
  \bibinfo{journal}{Phys. Rev. Lett.} \textbf{\bibinfo{volume}{80}},
  \bibinfo{pages}{4514} (\bibinfo{year}{1998}).

\bibitem[{\citenamefont{Albrecht et~al.}(1998)\citenamefont{Albrecht, Reining,
  {Del S}ole, and Onida}}]{albrecht98}
\bibinfo{author}{\bibfnamefont{S.}~\bibnamefont{Albrecht}},
  \bibinfo{author}{\bibfnamefont{L.}~\bibnamefont{Reining}},
  \bibinfo{author}{\bibfnamefont{R.}~\bibnamefont{{Del S}ole}},
  \bibnamefont{and} \bibinfo{author}{\bibfnamefont{G.}~\bibnamefont{Onida}},
 \bibinfo{journal}{Phys. Rev. Lett.} \textbf{\bibinfo{volume}{80}},
  \bibinfo{pages}{4510} (\bibinfo{year}{1998}).

\bibitem[{\citenamefont{Rohlfing and Louie}(1998)}]{rohlfing98}
\bibinfo{author}{\bibfnamefont{M.}~\bibnamefont{Rohlfing}} \bibnamefont{and}
  \bibinfo{author}{\bibfnamefont{S.~G.} \bibnamefont{Louie}},
  \bibinfo{journal}{Phys. Rev. Lett.} \textbf{\bibinfo{volume}{81}},
  \bibinfo{pages}{2312} (\bibinfo{year}{1998}).

\bibitem[{\citenamefont{Ruini et~al.}(2002)\citenamefont{Ruini, Caldas, Bussi,
  and Molinari}}]{ruini02}
\bibinfo{author}{\bibfnamefont{A.}~\bibnamefont{Ruini}},
  \bibinfo{author}{\bibfnamefont{M.~J.} \bibnamefont{Caldas}},
  \bibinfo{author}{\bibfnamefont{G.}~\bibnamefont{Bussi}}, \bibnamefont{and}
  \bibinfo{author}{\bibfnamefont{E.}~\bibnamefont{Molinari}},
  \bibinfo{journal}{Phys. Rev. Lett.} \textbf{\bibinfo{volume}{88}},
  \bibinfo{pages}{206403} (\bibinfo{year}{2002}).

\bibitem[{\citenamefont{Reich and Thomsen}(2000)}]{reich00c}
\bibinfo{author}{\bibfnamefont{S.}~\bibnamefont{Reich}} \bibnamefont{and}
  \bibinfo{author}{\bibfnamefont{C.}~\bibnamefont{Thomsen}},
  \bibinfo{journal}{Phys. Rev. B} \textbf{\bibinfo{volume}{62}},
  \bibinfo{pages}{4273} (\bibinfo{year}{2000}).

\end{thebibliography}

\end{document}